\documentclass[prl,twocolumn,amsmath,letterpaper]{revtex4}
\usepackage{amssymb}
\usepackage{graphicx}
\usepackage{array}
\usepackage{hhline}
\usepackage{longtable}

\renewcommand{\thetable}{\Roman{table}} \thetable

\begin{document}

\title{Field-Driven Hysteresis of the d=3 Ising Spin Glass: Hard-Spin Mean-Field Theory}
\author{Burcu Y\"{u}cesoy$^{1}$ and A. Nihat Berker$^{2-4}$}

\affiliation{$^1$Department of Physics, Istanbul Technical
University, Maslak 34450, Istanbul, Turkey,}

\affiliation{$^2$Department of Physics, Ko\c{c} University, Sar\i
yer 34450, Istanbul, Turkey,}

\affiliation{$^3$Feza G\"ursey Research Institute, T\"UB\.ITAK -
Bosphorus University, \c{C}engelk\"oy 34680, Istanbul, Turkey,}

\affiliation{$^4$Department of Physics, Massachusetts Institute of
Technology, Cambridge, Massachusetts 02139, U.S.A.}

\begin{abstract}
Hysteresis loops are obtained in the Ising spin-glass phase in
$d=3$, using frustration-conserving hard-spin mean-field theory. The
system is driven by a time-dependent random magnetic field $H_Q$
that is conjugate to the spin-glass order $Q$, yielding a
field-driven first-order phase transition through the spin-glass
phase.  The hysteresis loop area $A$ of the $Q-H_Q$ curve scales
with respect to the sweep rate $h$ of magnetic field as $A-A_{0}$
$\sim $ $h^{b}$. In the spin-glass and random-bond ferromagnetic
phases, the sweep-rate scaling exponent $b$ changes with temperature
$T$, but appears not to change with antiferromagnetic bond
concentration $p$. By contrast, in the pure ferromagnetic phase, $b$
does not depend on $T$ and has a sharply different value than in the
two other phases.

PACS numbers: 75.10.Nr, 75.60.Ej, 64.60.Ht, 05.70.Ln
\end{abstract}
\maketitle
\def\s{\rule{0in}{0.28in}}

\setlength{\LTcapwidth}{\columnwidth}

Frustration and non-equilibrium effects induce complicated ordering
behaviors that challenge the methods of statistical physics. Perhaps
the most ubiquitous non-equilibrium effect, hysteresis is the
current topic of intense fundamental and applied studies.  The
effects of a spatially uniform linearly driven magnetic field on the
ferromagnetic phase of the $n$-component vector model in $d = 3$
\cite{ZZS} and of the pure (no quenched randomness) Ising model in
$d = 2,3,4$ \cite{ZZX, ZZ} have been studied. In these works, the
scaling exponents of the hysteresis loop area with respect to the
sweep rate are obtained.  Spatially uniform pulsed, sinusoidally
oscillating, and stochastically varying magnetic fields on the
ferromagnetic phase of the pure Ising model in $d = 2,3$
\cite{Chakrabarti1, Chakrabarti2} and spatially uniform square-wave
magnetic fields on the ferromagnetic phase of the pure Ising model
in $d = 2$ \cite{Rikvold} have been studied. Two current detailed
experimental studies of hysteresis are in Refs.\cite{Bonnot,
Pierce}.

\begin{figure}[h]
\includegraphics*[scale=1]{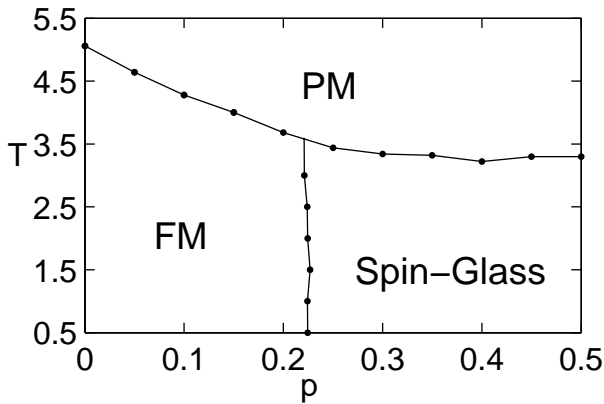}
\caption{Phase diagram from hard-spin mean-field theory for the d=3
Ising spin glass. All phase boundaries are second order.}
\label{fig2}
\end{figure}

\begin{figure}[h]
\includegraphics*[scale=0.96]{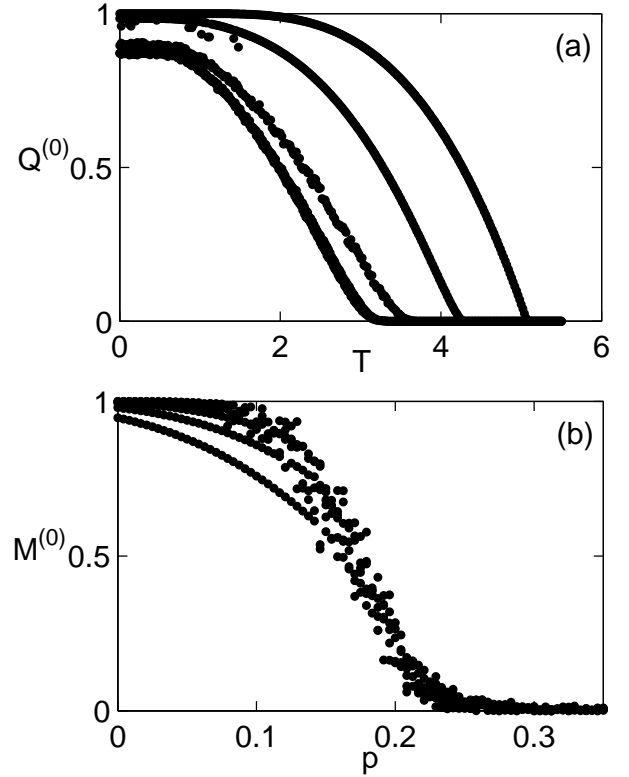}
\caption{(a) Equilibrium spin-glass order parameter $Q^{(0)}$ as a
function of temperature $T=J^{-1}$.  The curves, from top to bottom,
are for $p=0,0.1,0.2,0.3,0.5$.  The latter two curves overlap.  (b)
Equilibrium magnetization $M^{(0)}$ as a function of concentration
$p$.  The curves, from top to bottom, are for
$T=0.5,1.0,1.5,2.0,2.5,3.0$.}\label{fig1}
\end{figure}

\begin{figure}[h]
\includegraphics*[scale=0.96]{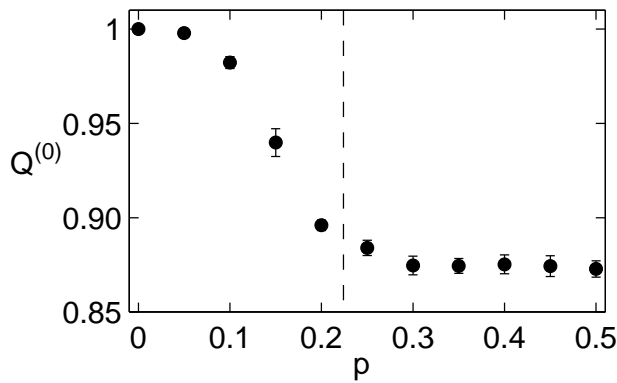}
\caption{Zero-temperature spin-glass order parameter $Q^{(0)}$ as a
function of antiferromagnetic bond concentration $p$, obtained by
averaging over 10 realizations, with the standard deviation being
used as the error bar. The dashed line indicates the transition
between the two phases, whose position is obtained from the phase
diagram in Fig.1.}\label{fig1a}
\end{figure}

\begin{figure}[h]
\centering
\includegraphics*[scale=1]{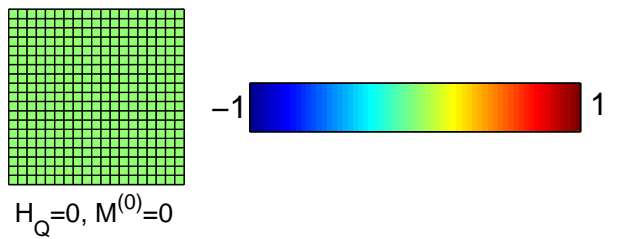}
\caption{The top-row figures are from a hysteresis loop in the
ferromagnetic phase with quenched random antiferromagnetic bonds,
$T=1.5, p=0.15, h=0.005$. The middle-row figures are from a
hysteresis loop in the spin-glass phase, $T=1.5, p=0.4, h=0.005$.
Left: calculated equilibrium local magnetizations $m_{i}^{(0)}$ in a
cross-section of the three-dimensional system.  A hysteresis loop is
started from these systems.  Middle: local magnetizations $m_{i}(t)$
at the first cancellation point, $M(t)=0$ (top row) and $Q(t)=0$
(middle row), of the first hysteresis loop.  Left: local
magnetizations at the first reversal point, $M(t)=-M(0)$ and
$Q(t)=-Q(0)$, which occurs when the first hysteresis half-loop is
completed. The bottom cross-section shows the vanishing equilibrium
local magnetizations everywhere in the paramagnetic phase, to be
contrasted with the spin-glass cross-section immediately above it:
the global magnetization $M^{(0)}=0$ in both cases.} \label{fig3}
\end{figure}

In the present study, hard-spin mean-field theory, developed
specifically to respect frustration \cite{Netz1, Netz3}, is used to
study the non-equilibrium behavior of the field-driven first-order
phase transition that is implicit, but to-date unstudied, in
spin-glass ordering. For the Ising spin-glass on a cubic lattice,
the phase diagram is obtained and the temperature- and
concentration-dependent ordering of the spin-glass phase is
microscopically determined.  The random magnetic field that is
conjugate to this microscopic order is then identified and used to
induce a first-order transition and hysteresis loops. We find
qualitatively and quantitatively contrasting scaling behaviors in
spin-glass, quenched random-bond ferromagnetic, and pure
ferromagnetic phases of the system.

The model is defined by the Hamiltonian

\begin{equation}\label{eq:1}
-\beta H=\underset{<ij>}{\sum }J_{ij}s_{i}s_{j}+\underset{i}{\sum }%
H_{i}(t)s_{i}\text{,}
\end{equation}%

\noindent where $s_{i}=\pm 1$ at each site $i$ of a cubic lattice
and $<ij>$ denotes summation over nearest-neighbor pairs. The bond
strengths $J_{ij}$ are equal to $-J$ with quenched probability $p$
and $+J$ with probability $1-p$, respectively corresponding to
antiferromagnetic and ferromagnetic coupling. $H_{i}(t)$ is a
linearly swept quenched random magnetic field, itself determined, as
explained below, by the spin-glass local order of this system.

\begin{figure}[h]
\centering
\includegraphics*[scale=1]{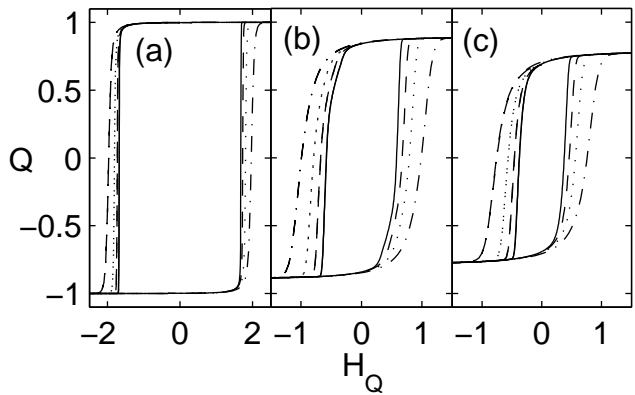}
\caption{Hysteresis loops for different values of the sweep rate $h$
for (a) the pure ferromagnetic phase, $p=0$, (b) the ferromagnetic
phase with quenched random antiferromagnetic bonds, $p=0.15$, (c)
the spin-glass phase, $p=0.4$, all at $T=1.5$.  The loops are, from
outer to inner, for sweep rates
$h=0.05,0.02,0.01,0.005$.}\label{fig4}
\end{figure}

For our calculations we use the hard-spin mean-field theory
\cite{Netz1, Netz3, Banavar, Netz2, Netz4, Netz5, Berker,
Kabakcioglu2, McKayAbs, Akguc, McKay2, Monroe, Pelizzola,
Kabakcioglu, Kaya}, a method which is nearly as simply implemented
as the conventional mean-field theory but which conserves
frustration by incorporating the effect of the full magnitude of
each spin. The self-consistent equation for local magnetizations
$m_{i}$ in hard-spin mean-field theory is
\begin{equation}\label{eq:2}
m_{i}=\underset{\{s_{j}\}}{\sum }\left[ \underset{j}{\prod }P\left(
m_{j},s_{j}\right) \right] \tanh \left( \underset{j}{\sum }
J_{ij}s_{j}+H_{i}(t)\right) \text{,}
\end{equation}

\noindent where the sum $\{s_{j}\}$ is over all interacting neighbor
configurations and the sum and the product over $j$ are over all
sites that are coupled to site $i$ by interaction $J_{ij}$. The
single-site probability distribution $ P(m_{j},s_{j})$ is
$(1+m_{j}s_{j})/2$.  The hard-spin mean-field theory has been used
in time-dependent systems, in the study of field-cooled and
zero-field cooled magnetizations in spin glasses.\cite{McKayAbs}

\noindent\textit{Equilibrium Phase Diagram} - The equilibrium local
magnetizations $m_{i}^{(0)}$ are determined by simultaneously
solving $N$ coupled Eqs.(\ref{eq:2}) for all $N$ sites $i$ of the
system, at zero external magnetic field, $H=0$.  For $0<p<1$, the
system is degenerate, and many local magnetization solutions exist
and are reached by hard-spin mean-field theory.  The phase diagram
(Fig.\ref{fig2}) is obtained from temperature $T=J^{-1}$ and
concentration $p$ scans of the equilibrium spin-glass order
parameter $Q^{(0)} = \frac{1}{N}\Sigma _i m_i^2$ and magnetization
$M^{(0)} = \frac{1}{N}\Sigma _i m_i$, illustrated in Fig.\ref{fig1},
obtained by averaging over 20 realizations for a $N=20^{3}$ spin
system. The results do not change if a larger system is used. In the
resulting phase diagram shown in Fig.\ref{fig2}, the transition
temperatures are gauged by comparing $T_C$ at $p=0$: The precise
value \cite{Landau} is 4.51, the ordinary mean-field value is 6, the
value obtained here is 5.06~. Thus, the transition temperatures are
exaggerated as expected from a mean-field theory, but considerably
improved over ordinary mean-field theory. Our obtained transition
concentrations between the ferromagnetic and spin-glass phases are
$p=0.22$, in excellent agreement with the precise value of $p=0.23$
\cite{Ozeki}.

Fig.\ref{fig1a} shows the zero-temperature spin-glass order
parameter $Q^{(0)}$ as a function of antiferromagnetic bond
concentration $p$. It seen that, as soon as frustration is
introduced via the antiferromagnetic bonds, order does not saturate
at zero temperature, both in the ferromagnetic and spin-glass
phases, the latter of course showing more unsaturation.  Moreover,
the left column of Fig.\ref{fig3} shows the equilibrium local
magnetizations $m_i$ in a cross-section of the system, in the
ferromagnetic and spin-glass phases.  These magnetization
cross-sections are remarkably similar to the renormalization-group
results \cite{Yesilleten} and are consistent with the chaotic
rescaling picture of the spin-glass phase \cite{McKay}.

\begin{figure}[h]
\centering
\includegraphics*[scale=1]{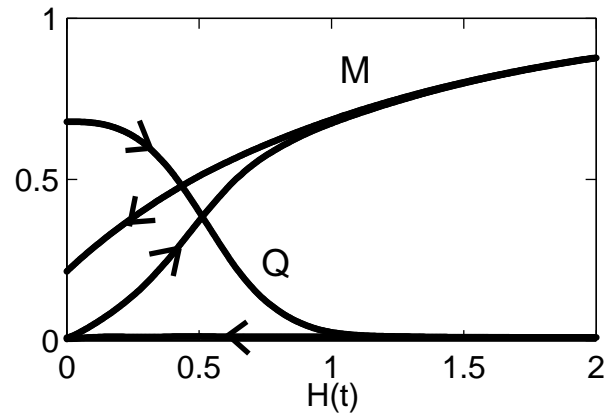}
\caption{ Spin-glass order parameter $Q(t)$ and uniform
magnetization $M(t)$ curves obtained when, in the spin-glass phase,
the uniform magnetic field $H(t)$ is turned on and then off with
sweep rate $h=0.005$.  In this figure, $p=0.4, T=1.5$.}\label{fig5}
\end{figure}

\noindent\textit{Spin-Glass Hysteresis Loops} - The quenched random
magnetic field that is conjugate to the microscopic order is $H_i(t)
= H_Q(t)m_i^{(0)}$ in Eq.(\ref{eq:1}), where the $m_i^{(0)}$ are the
equilibrium local magnetizations obtained with Eq.(\ref{eq:2}) for a
given $T, p$ . Hysteresis loops in the spin-glass order
$Q(t)=\frac{1}{N} \Sigma_i m_{i}(t)m_{i}^{(0)}$ are obtained in the
ordered phases, spin-glass or ferromagnetic, by cycling $H_Q(t)$ at
constant $T, p$, via a step of magnitude $h$ for each time unit.
Thus, at time $t=0$, $Q(t=0) = Q^{(0)}$, the equilibrium spin-glass
order parameter.  A time unit is $N$ updatings of Eq.(\ref{eq:2}) at
randomly selected sites. Thus, $h$ is the sweep rate of the linearly
driven \cite{ZZS, ZZX, ZZ} magnetic field.  For comparison, it takes
of the order of $40 N$ random updatings to establish equilibrium in
the pure ferromagnetic phase and $2000 N$ random updatings to
estabish equilibrium in the spin-glass phase.  Thus, during the
sweeps, each time step with $N$ updatings systematically moves in
the direction of the changed external field but, within the loops,
does not catch up with equilibrium. The resulting hysteresis curves
are illustrated in Figs.\ref{fig4}. After one cycling, the
subsequent hysteresis loops for a given sweep rate coincide, and are
shown in Figs.\ref{fig4} and used in the scaling analysis further
below.

\noindent\textit{Cycling Effect of a Uniform Magnetic Field on
Spin-Glass Order} - As a contrast to the hysteretic effect of the
conjugate quenched random magnetic field $H_Q(t)$ introduced above,
Fig.\ref{fig5} shows the effect on the spin-glass phase of turning
on and then off a uniform magnetic field $H(t)$ at a sweep rate $h$.
As expected, the spin-glass order $Q(t)$ starts at a finite value
and returns to zero, while the uniform magnetization
$M(t)=\frac{1}{N}\Sigma_i m_{i}(t)$ starts at zero and returns to a
finite value.

\begin{figure}[h]
\includegraphics*[scale=1]{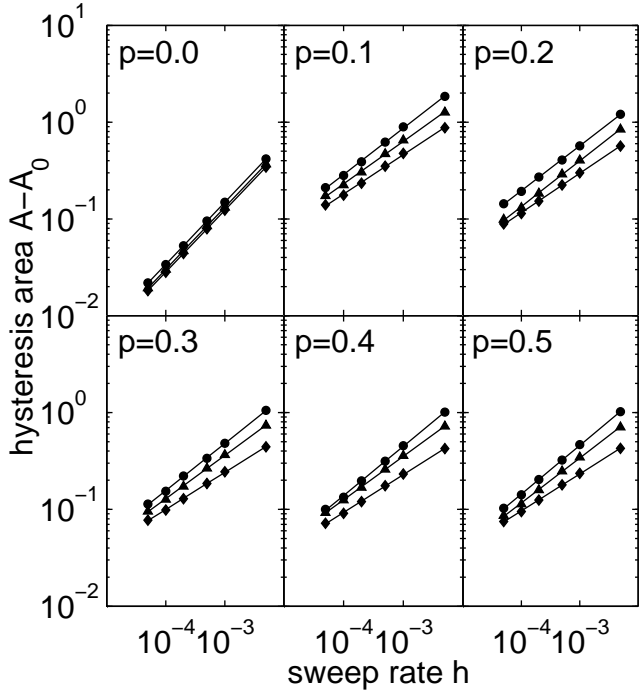}
\caption{The hysteresis area $A-A_{0}$ versus sweep rate $h$ scaling
curves for $T=1.0 (\bullet),1.5 (\blacktriangle),2.0(\blacklozenge)$
at different concentrations $p$.}\label{fig6}
\end{figure}

\begin{figure}[h]
\includegraphics*[scale=1]{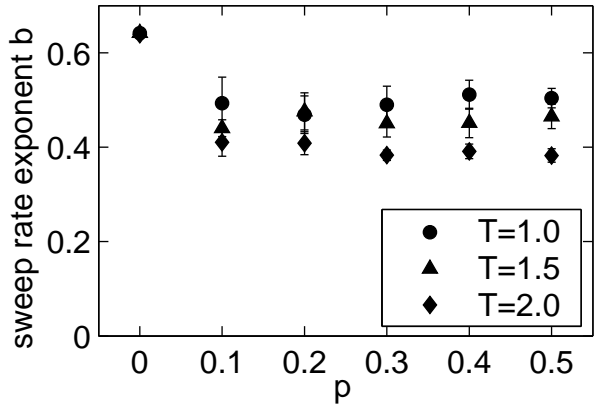}
\caption{The sweep-rate scaling exponent $b$ versus concentration
$p$ for $T=1.0,1.5,2.0$.  These results are obtained by averaging
over 10 realizations, with the standard deviation being used as the
error bar.}\label{fig7}
\end{figure}

\noindent\textit{Spin-Glass Hysteresis Area Scaling} - The energy
dissipation of a first-order phase transition is obtained from the
hysteresis area $A$\ of the $Q-H_Q$ curve: $A=\oint QdH_Q$.  At
fixed $T,p$, the loop area $A$ decreases with decreasing sweep rate
$h$ and finally reaches a value of $A_{0}$. The area can be scaled
as $A=A_{0}+f(T)h^{b}$.\cite{ZZ}  The $(A-A_{0})$ versus sweep rate
$h$ scaling curves are shown in Figs.\ref{fig6}, for the pure
ferromagnetic, quenched random-bond ferromagnetic, and spin-glass
phases for various temperatures, where $A_0$ is fitted. The
resulting sweep-rate exponents $b$ are given in Fig.\ref{fig7} and
Table I.  From these results, we deduce that in the pure
ferromagnetic phase, $p=0$, the exponent $b$ is independent of
temperature, as found previously \cite{ZZ}.  However, the value of
$b=0.64$ that we find here, under hard-spin mean-field dynamics, is
distinctly different from that of $b=2/3$ found in Ref.\cite{ZZ}
under ordinary mean-field dynamics, thereby constituting a different
dynamic universality class.  By contrast, in the quenched
random-bond ferromagnetic phase and in the spin-glass phase, the
value of $b$ is distinctly smaller than that in the pure
ferromagnetic phase, and dependent on temperature. Across both of
these two phases, there appears to be no dependence of $b$ on
concentration.

\squeezetable
\begin{table}[t]
\renewcommand{\arraystretch}{1}
\begin{tabular}{|l|l|l|l|l|l|l|}
\hline
$~T $ & $p=0$ & $~p=0.1$ & $~p=0.2$ & $~p=0.3$ & $~p=0.4$ & $~p=0.5$\\
\hline
$1.0$ & ~0.64 &0.49$\pm 0.06$ &0.47$\pm0.04$ &0.49$\pm0.04$ &0.51$\pm0.03$ &0.50$\pm0.02$ \\
\hline
$1.5$ & ~0.64 &0.44$\pm 0.02$ &0.48$\pm0.04$ &0.45$\pm0.03$ &0.45$\pm0.03$ &0.47$\pm0.03$ \\
\hline
$2.0$ & ~0.64 &0.41$\pm 0.03$ &0.41$\pm0.02$ &0.38$\pm0.01$ &0.39$\pm0.02$ &0.38$\pm0.01$ \\
\hline
\end{tabular}\caption{The sweep-rate scaling exponents $b$ at different
temperatures and concentrations in the ferromagnetic and spin-glass
phases.} \label{tab1}\end{table}

\begin{acknowledgments}
We thank Y. \"Oner for drawing our attention to hysteresis problems
in frustrated systems and for useful discussions.  We thank M.
Hinczewski, A. Kabak\c{c}\i o\u{g}lu, and M.C. Yalab\i k for useful
discussions. This research was supported by the Scientific and
Technical Research Council (T\"UB\.ITAK) and by the Academy of
Sciences of Turkey.
\end{acknowledgments}


\begin{thebibliography}{}

\bibitem{ZZS} F. Zhong, J.X. Zhang, and G.G. Siu, J. Phys.: Cond. Matt. {\bf 6}, 7785 (1994).
\bibitem{ZZX} F. Zhong, J.X. Zhang, and L. Xiao, Phys. Rev. E {\bf 52}, 1399 (1995).
\bibitem{ZZ} G.P. Zheng and J.X. Zhang, J. Phys.: Cond. Matt. {\bf 10}, 1863 (1998).
\bibitem{Chakrabarti1} B.K. Chakrabarti and M. Acharyya, Rev. Mod. Phys. {\bf 71}, 847
(1999).
\bibitem{Chakrabarti2} A. Chatterjee and B.K. Chakrabarti, Phase Tr. {\bf 77}, 581 (2004).
\bibitem{Pierce} M.S. Pierce, C.R. Buechler, L.B. Sorensen, S.D. Kevan, E.A. Jagla,
J.M. Deutsch, T. Mai, O. Narayan, J.E. Davies, K. Liu, G.T. Zimanyi,
H.G. Katzgraber, O. Hellwig, E.E. Fullerton, P. Fischer, and J.B.
Kortright, Phys. Rev. B {\bf 75}, 144406 (2007).
\bibitem{Bonnot} E. Bonnot, R. Romero, X. Illa, L. Manosa, A. Planes, and E. Vives,
Phys. Rev. B {\bf 76}, 064105 (2007).
\bibitem{Rikvold} D.T. Robb, P.A. Rikvold, A. Berger, and M.A. Novotny, Phys. Rev. E {\bf 76}, 021124 (2007).
\bibitem{Netz1} R.R. Netz and A.N. Berker, Phys. Rev. Lett. {\bf 66}, 377 (1991).
\bibitem{Netz3} R.R. Netz and A.N. Berker, J. Appl. Phys. {\bf 70}, 6074 (1991).
\bibitem{Banavar} J.R. Banavar, M. Cieplak, and A. Maritan, Phys. Rev. Lett. {\bf 67}, 1807 (1991).
\bibitem{Netz2} R.R. Netz and A.N. Berker, Phys. Rev. Lett. {\bf 67}, 1808 (1991).
\bibitem{Netz4} R.R. Netz, Phys. Rev. B {\bf 46}, 1209 (1992).
\bibitem{Netz5} R.R. Netz, Phys. Rev. B {\bf 48}, 16113 (1993).
\bibitem{Berker} A.N. Berker, A. Kabak\c{c}\i o\u{g}lu, R.R. Netz, and M.C. Yalab\i k, Turk.
J. Phys. {\bf 18}, 354 (1994).
\bibitem{Kabakcioglu2} A. Kabak\c{c}\i o\u{g}lu, A.N. Berker, and M.C. Yalab\i k, Phys. Rev. E {\bf 49}, 2680
(1994).
\bibitem{McKayAbs} E.A. Ames and S.R. McKay, J. Appl. Phys. {\bf 76}, 6197 (1994).
\bibitem{Akguc} G.B. Akg\"u\c{c} and M.C. Yalab\i k, Phys. Rev. E {\bf 51}, 2636 (1995).
\bibitem{McKay2} J.E. Tesiero and S.R. McKay, J. Appl. Phys. {\bf 79}, 6146,
(1996).
\bibitem{Monroe} J.L. Monroe, Phys. Lett. A {\bf 230}, 111 (1997).
\bibitem{Pelizzola} A. Pelizzola and M. Pretti, Phys. Rev. B {\bf 60}, 10134 (1999).
\bibitem{Kabakcioglu} A. Kabak\c{c}\i o\u{g}lu, Phys. Rev. E {\bf 61}, 3366 (2000).
\bibitem{Kaya} H. Kaya and A.N. Berker, Phys. Rev. E {\bf 62}, R1469 (2000).
\bibitem{Landau} A.M. Ferrenberg and D.P. Landau, Phys. Rev. B {\bf 44}, 5081 (1991).
\bibitem{Ozeki} Y. Ozeki and N. Ito, J. Phys. A {\bf 31}, 5451 (1998).
\bibitem{Yesilleten} D. Ye\c{s}illeten and A.N. Berker, Phys. Rev. Lett. {\bf 78}, 1564 (1997).
\bibitem{McKay} S.R. McKay, A.N. Berker, and S. Kirkpatrick, Phys. Rev. Lett. {\bf 48}, 767 (1982).
\end{thebibliography}
\end{document}